\documentclass[3p,times,procedia]{elsarticle}
\pdfoutput=1
\usepackage{ecrc}
\usepackage[bookmarks=false]{hyperref}
    \hypersetup{colorlinks,
      linkcolor=blue,
      citecolor=blue,
      urlcolor=blue}

\usepackage{booktabs} 

\usepackage{arydshln}
\usepackage{url}
\usepackage{xspace}
\usepackage{enumitem}

\newcommand{\cx}[0]{\multicolumn{1}{c}{\textbullet}}

\newcommand{\privmod}{PM-MoDaC\xspace}

\usepackage[strings]{underscore} 
\usepackage{wrapfig}

\volume{00}

\firstpage{1}

\journalname{Procedia Computer Science}

\runauth{Felix Beierle et al.}

\jid{procs}

\usepackage{amssymb}

\usepackage[figuresright]{rotating}

\begin{document}
\begin{frontmatter}

\dochead{The 15th International Conference on Mobile Systems and Pervasive Computing \\ (MobiSPC 2018)}%

\title{Context Data Categories and Privacy Model\\for Mobile Data Collection Apps}

\author[a]{Felix Beierle\corref{cor1}}
\author[a]{Vinh Thuy Tran}
\author[b]{Mathias Allemand}
\author[c]{Patrick Neff}
\author[c]{Winfried Schlee}
\author[d]{Thomas Probst}
\author[e]{R\"udiger Pryss}
\author[f]{Johannes Zimmermann}

\address[a]{Service-centric Networking, Telekom Innovation Laboratories, Technische Universit\"at Berlin, Berlin, Germany}
\address[b]{Department of Psychology, University of Zurich, Zurich, Switzerland}
\address[c]{Clinic and Policlinic for Psychiatry and Psychotherapy, University of Regensburg, Regensburg, Germany}
\address[d]{Department for Psychotherapy and Biopsychosocial Health, Danube University Krems, Krems, Austria}
\address[e]{Institute of Databases and Information Systems, Ulm University, Ulm, Germany}
\address[f]{Psychologische Hochschule Berlin, Berlin, Germany}

\begin{abstract}
Context-aware applications stemming from diverse fields like mobile health,
recommender systems, and mobile commerce potentially benefit from knowing
aspects of the user's personality.
As filling out personality questionnaires is tedious,
we propose the prediction of the user's personality from
smartphone sensor and usage data.
In order to collect data for researching the relationship
between smartphone data and personality, we developed
the Android app TYDR (Track Your Daily Routine) which
tracks smartphone data
and utilizes psychometric personality questionnaires.
With TYDR,
we track a larger variety of smartphone data than similar existing apps,
including metadata on notifications, photos taken, and music played back by the user.
For the development of TYDR, we introduce a
general context data model consisting of four categories
that focus on the user's different types of interactions
with the smartphone:
\emph{physical} conditions and activity, \emph{device} status and usage, \emph{core functions} usage, and \emph{app} usage.
On top of this, we develop the privacy model \privmod specifically for apps
related to the collection of mobile data, consisting of
nine proposed privacy measures.
We present the implementation of all of those measures in TYDR.
Although the utilization of the user's personality based on the usage of his or her smartphone is a challenging endeavor, it seems to be a promising approach for various types of context-aware mobile applications.

\end{abstract}

\begin{keyword}
ubiquitous computing; context-aware computing; psychometrics; sensor data

\end{keyword}
\cortext[cor1]{Corresponding author. Tel.: +49 30 8353 54265.}
\end{frontmatter}

\email{beierle@tu-berlin.de}

\section{Introduction}
\label{sec:intro}
The modern smartphone is a small personal computer
that is used for a large variety of tasks in different contexts.
A multitude of sensors and an omnipresent internet connectivity
make apps aware of the user's context.
This context can be used to personalize or contextualize
applications, for example, by recommending something based
on the current time and location.
Oftentimes, the context that is taken into consideration
is limited to directly measurable factors like
location, battery status, or installed apps.

Having additional context data about the user's personality
could improve context-aware systems from different domains, e.g.,
mobile health, personalization and recommendations, or mobile commerce.
Mobile health applications could benefit from
personality data
for the diagnosis or treatment of patients
(e.g., \cite{PryssMobileCrowdSensing2015,MichaelJ.RocheEnrichingPsychologicalAssessment2014,ZimmermannIntegratingstructuredynamics,PryssProspectivecrowdsensingretrospective2018}).
Context-aware recommender systems may benefit from personality data,
as was shown in a recent study with the MovieLens recommender system
\cite{KarumurPersonalityUserPreferences2017}.
The importance of personality for the attitude towards advertising and mobile commerce
is highlighted for example in \cite{Myersmoderatingeffectpersonality2010, ZhouEffectsPersonalityTraits2011}.

Psychological research suggests that there are links between personality
traits and everyday preferences \cite{beierle-towards-2017}.
With the smartphone, we will be able to track different types of data that might
reflect the user's personality:
the smartphone's sensors can track the user's physical context
and the operating system can track the user's interaction with the smartphone and its apps.
We argue that, after collecting data labeled with the personality of the user, we might
be able to predict (aspects of) the user's personality from sensor and usage data
without applying questionnaires.

In order to collect data to perform a study analyzing the relationship between
smartphone data and personality, we developed
the Android app TYDR (Track Your Daily Routine).
TYDR collects smartphone sensor and usage data
as well as applies standardized psychological questionnaires to the user.
In \cite{BeierleMobileSoft2018},
we highlighted some aspects about the development process of the app,
relating to the implementation of sensor data collection
and some privacy aspects.
In this paper, we focus on two aspects that researchers face
when planning similar studies and designing applications like TYDR.
The first aspect is the development of a general model of context data
for smartphone applications that takes into account the user's interaction with the phone.
As context data like detailed location information or app usage statistics
is highly sensitive, the second aspect we focus on is introducing a privacy model,
specifically for smartphone apps relating to mobile data collection.
The main contributions of this paper thus are:
\begin{itemize}[noitemsep, topsep=2pt]
	\item We propose a general context data model for smartphone applications.
	\item We introduce the privacy model \privmod for apps relating to mobile data collection.
	\item We give an overview of the implementation of the introduced privacy model in the Android app TYDR.
\end{itemize}

The remainder of this paper is structured as follows. In Section \ref{sec:rel-work},
we review related work,
showing that none of the existing projects take into account all the available
data sources present on current smartphones.
We present a general model of context data for smartphone applications
in Section \ref{sec:cat-context} and
introduce our privacy model \privmod in Section \ref{sec:privacy}.
We show the implementation of \privmod in TYDR in Section \ref{sec:sys-arch}
before concluding
in Section \ref{sec:concl-fw}.

\section{Related Work and Study Planning}
\label{sec:rel-work}

In Table \ref{tbl:rel-work}, we give an overview of related studies
that correlated sensor and/or smartphone usage data
with user information related to personality.
Some of those studies have been conducted with feature phones, before the advent of smartphones
\cite{ChittaranjanWhoWhoBigFive2011, ChittaranjanMininglargescalesmartphone2013, ButtPersonalityselfreported2008}.
The \emph{data sources} given in the table
differ in their level.
For example, accelerometer data is low level sensor data,
while the current activity (walking, in car, etc.) or a daily step count
is higher level sensor data that utilizes accelerometer data.
The available data sources depend on the used OS and on the
available libraries and SDKs.
In the table, we list the sources
mentioned in the cited papers.
There might be some steps in between low level sensor data and the user's personality, like
estimating the user's sleep pattern utilizing low level sensor data like phone un-/lock events. For overviews related to determining
higher level features from lower level sensor data see \cite{HarariSmartphonesensingmethods2017,MohrPersonalSensingUnderstanding2017}.

\begin{table}[t]
	\small
	\centering
	\caption{Overview of data sources and user information that were correlated in previous studies.}
	\label{tbl:rel-work}
	\begin{tabular}{@{}llll@{}}
		\toprule
		\textbf{Data Sources}                                                       & \textbf{User Information}     & \textbf{Property} & \textbf{References}             						\\ 
		Bluetooth, calls, sms, calling profiles, application usage (pre-smartphone) & personality traits            & static & \cite{ChittaranjanWhoWhoBigFive2011, ChittaranjanMininglargescalesmartphone2013}                  																										\\ \hdashline[0.5pt/2.5pt]
		calls, sms, changing ringtones and wallpapers (pre-smartphone)              & personality traits            & static & \cite{ButtPersonalityselfreported2008}                \\ \hdashline[0.5pt/2.5pt]
		location                                                                    & personality traits            & static & \cite{ChorleyPersonalitylocationbasedsocial2015}  	\\ \hdashline[0.5pt/2.5pt]
		technology usage times                                                      & personality traits            & static & \cite{GroverDigitalFootprintsPredicting2017}      	\\ \hdashline[0.5pt/2.5pt]
		calls, sms, location					                                    & personality traits            & static & \cite{deMontjoyePredictingPersonalityUsing2013}      \\ \hdashline[0.5pt/2.5pt]
		installed apps                                                              & personality traits     	    & static & \cite{XuUnderstandingimpactpersonality2016}  		\\ \hdashline[0.5pt/2.5pt]
		app usage																	& personality traits   			& static & \cite{StachlPersonalityTraitsPredict2017}            \\ \hdashline[0.5pt/2.5pt]
		calls, sms, proximity data, weather                                         & daily stress                  & dynamic & \cite{BogomolovDailyStressRecognition2014}            \\ \hdashline[0.5pt/2.5pt]
		accelerometer, Bluetooth, location											& emotions						& dynamic & \cite{RachuriEmotionSenseMobilePhones2010}			 \\	\hdashline[0.5pt/2.5pt]
		location                                                                    & depressive states             & dynamic & \cite{CanzianTrajectoriesDepressionUnobtrusive2015}   \\ \hdashline[0.5pt/2.5pt]
		email, sms, calls, websites, location, app usage				            & mood                          & dynamic & \cite{LiKamWaMoodScopeBuildingMood2013}               \\ 		\bottomrule
	\end{tabular}
\end{table}

The ground truth for \emph{user information} is typically assessed via self-report methods, i.e., questionnaires.
Often, the authors of the studies describe use cases to illustrate what the predicted user
information could be meaningful for.
Most of the studies aim at use cases related to mobile health
or context-aware recommender systems,
e.g., recommending new apps based on the personality correlated with
already installed apps \cite{XuUnderstandingimpactpersonality2016}.
Some studies go further than correlating data with the personality of the user.
The StudentLife project, for example, collected sensor data and queried student participants with a
variety of questionnaires to predict
mental health and academic performance \cite{WangStudentLifeAssessingMental2014, WangSmartGPAHowSmartphones2015, WangStudentLifeUsingSmartphones2017}.

In the \emph{property} column of Table \ref{tbl:rel-work},
we distinguish related studies
as
being \emph{static} or \emph{dynamic}.
A static system will look for
information such as personalty traits that are relatively stable.
A dynamic study will try to find correlations between sensor/usage data
and changing aspects about the user, for example, mood or stress level \cite{LiKamWaMoodScopeBuildingMood2013,BogomolovDailyStressRecognition2014}.

There are some additional projects that are related to our research.
Sensus \cite{XiongSensusCrossplatformGeneralpurpose2016}, LiveLabs \cite{JayarajahLiveLabsBuildingInSitu2016},
and AWARE \cite{FerreiraAWAREMobileContext2015} aim at providing researchers with
frameworks for conducting research related to collected sensor/smartphone usage data.
As far as the papers and website indicate, none of these frameworks provide
support for collecting music and photo metadata, which we enable with TYDR.

Most of the cited studies are interested in personality traits of the user.
The most prominent structural model of individual differences in personality traits is the
Big Five model \cite{mccrae-introduction-1992}, consisting of the trait domains
\emph{openness to experience},
\emph{conscientiousness},
\emph{extraversion},
\emph{agreeableness}, and
\emph{neuroticism}.
Moreover, the expression of personality traits fluctuates within persons across time \cite{FleesonStructureandProcessIntegratedView2001}.
For example, a person who scores high on neuroticism will experience negative
mood more often than other people, but may still vary considerably in the experience of negative
mood across time, e.g., depending on situational circumstances.
This within-person variability of emotions and behaviors is captured by the term "personality states."
In order to register those aspects, we utilize the PDD (Personality Dynamics Diary) questionnaire,
which captures the user's experience of daily situations and behaviors \cite{ZimmermannIntegratingstructuredynamics}.
With the results of a study with TYDR, we will investigate to what extend we can make daily predictions
about personality states based on context data.

\section{Categorization of Context Data}
\label{sec:cat-context}

\begin{wraptable}{R}{0.575\textwidth}
	\small
	\centering
	\caption{Context data model for the categorization of context data for smartphone applications. Perm. indicates if an explicit user permission is required (Android).}
	\label{tbl:context}
	\begin{tabular}{@{}llllll@{}}
\toprule
& \multicolumn{4}{c}{\textbf{Category}} & \textbf{Perm.} \\
\cmidrule(lr){2-5}

\textbf{}        	& Physical			& Device		& Core fct.	    & Apps			& 						\\
location			& \cx 				&				&				& 				& \cx					\\ \hdashline[0.5pt/2.5pt]
weather				& \cx				&				&				& 				& \multicolumn{1}{c}{(\textbullet)}					\\ \hdashline[0.5pt/2.5pt]
ambient light sensor& \cx				&				&				& 				& 						\\ \hdashline[0.5pt/2.5pt]
ambient noise level & \cx				&				&				& 				& \cx						\\ \hdashline[0.5pt/2.5pt]
accelerometer		& \cx				&				&				& 				& 						\\ \hdashline[0.5pt/2.5pt]
activity			& \cx 				&				&				& 				& 						\\ \hdashline[0.5pt/2.5pt]
steps				& \cx				&				&				& 				& 						\\ \hdashline[0.5pt/2.5pt]
phone un-/lock		&					& \cx			&				&				& 						\\ \hdashline[0.5pt/2.5pt]
headphone un-/plug	&	 				& \cx			&				& 				& 						\\ \hdashline[0.5pt/2.5pt]
battery and charging&					& \cx			&				& 				& 						\\ \hdashline[0.5pt/2.5pt]
Wifi				&					& \cx			&				& 				& 						\\ \hdashline[0.5pt/2.5pt]
Bluetooth			&					& \cx			&				& 				& 						\\ \hdashline[0.5pt/2.5pt]
calls metadata		&					&				& \cx			& 				& \cx					\\ \hdashline[0.5pt/2.5pt]
music metadata		&					&				& \cx			& 				& \multicolumn{1}{c}{(\textbullet)}						\\ \hdashline[0.5pt/2.5pt]
photos metadata		&					& 				& \cx			&				& \cx					\\ \hdashline[0.5pt/2.5pt]
notifications metadata&					& 				& \cx			& \cx			& \cx					\\ \hdashline[0.5pt/2.5pt]
app usage			&					&				&				& \cx			& \cx					\\ \hdashline[0.5pt/2.5pt]
app traffic			&					&				&				& \cx			& \cx					\\ \bottomrule
	\end{tabular}
\end{wraptable}

In broad terms, Dey defines context as something which is relevant to an
application \cite{dey-understanding-2001}.
Often, context is categorized into
\emph{device}, \emph{user}, \emph{physical} surrounding and activity, and \emph{temporal} aspects
\cite{yurur-context-awareness-2014}.
However, this does not reflect the users' interaction with the smartphone.
For example, the number of pictures taken or which apps a user is using
may yield important information about his/her context.
A user taking many pictures and using map applications might be at an unfamiliar place
that he/she enjoys.

In Table \ref{tbl:context}, we introduce
a general context data model for the categorization of context data for smartphone applications.
The four categories are
\emph{physical} conditions and activity, \emph{device} status and usage, \emph{core functions} usage, and \emph{app} usage.
Furthermore, an additional technical category constitutes the explicit permission
by the user in order to allow an app to access data from the given source.
This has important implications, e.g., for answering the question if it is possible to develop
a library for personality prediction that does not require explicit permissions.

\emph{\textbf{Physical}} conditions and activity deal with the physical context of the user
that is not related to the interaction with the smartphone.
Here, sensors deliver data without the user interacting with the phone,
e.g., location or taken steps.
The ambient light sensor typically offers data only when the screen is active,
so when the user is interacting with the phone.
However, as its data is related to the physical context, i.e.,
the light level of the environment of the user,
we regard it as part of the \emph{physical} category.

The category \emph{\textbf{device}} status and usage designates data that
is related to the status and the connectivity of the smartphone.
This comprises lock state, headphone connection status, battery level
and charging status as well as Wifi and Bluetooth connectivity.

\emph{\textbf{Core functions}} usage deals with the users' interaction
with core functionalities of the phone, regardless of which specific
apps they are using for it.
The core functions comprise calling, music listening, taking photos, and
dealing with notifications.

The fourth category is \emph{\textbf{app}} usage, dealing with
data about the usage and traffic of specific apps.
Notifications fit both in the \emph{core functions} and
the \emph{apps} categories because they can be related to either.

The \emph{permission} column
is based on the permission system introduced with
Android 6.0 (API 23).
Weather is given in parenthesis because it can only be collected if the location is available,
so it is bound to the location permission.
Music is given in parenthesis as well.
Most major music player apps or music streaming apps
automatically broadcast metadata about music that the user is currently listening to.
The broadcast events can be received by any app that subscribes as a
listener \cite{beierle-privacy-aware-2016}.
However, for Spotify, such broadcasting has to be activated manually.

In general, our context data model can be helpful for the development of any
context-aware service, e.g., in the areas of ubiquitous computing and mobile social networking \cite{beierle-towards-2015, BeierleTrustCom2018}.
After collecting data, we have to analyze to what extend
the quality of the context data varies between the variety of different
available Android devices.
Our context data categorization
allows to address different specific questions
based on our research question regarding the prediction of the user's personality.
Specific questions are, for example,
whether the physical context alone can predict personality,
how meaningful metadata is,
or how accurate the prediction can be if the user did not give any explicit permissions.

\section{\privmod{} -- Privacy Model for Mobile Data Collection Applications}
\label{sec:privacy}

As we are dealing with highly sensitive data, privacy concerns should have a high priority.
In this section, we present a comprehensive overview of measures that can be
taken to protect user privacy.
To the best of our knowledge, we are the first to provide such a comprehensive privacy model
for applications related to mobile data collection.

Of the reviewed related work, only one paper provides some details
about the processes and measures taken to ensure user privacy \cite{Kiukkonenrichmobilephone2010}.
Some works do not give any technical details about privacy protection
\cite{CanzianTrajectoriesDepressionUnobtrusive2015,BogomolovDailyStressRecognition2014}
or openly state that they disregarded the issue, e.g., \cite{RachuriEmotionSenseMobilePhones2010}:
"privacy is not a major concern for this system, since all users voluntarily agree to carry the devices for constant monitoring."
If there is information given about privacy protection, it is typically not very detailed
and usually only covers some of the aspects given in the following privacy model.

Our \textbf{Privacy Model for Mobile Data Collection Applications (\privmod)} comprises the following
nine privacy mechanisms (PM):

\paragraph{\textbf{(A) User Consent}}
Before installing the app, the user should be explained what data exactly
is being collected and for what purpose.
These are typical aspects covered in a privacy policy that the user has
to agree to before using an app.
The aspect of \emph{user consent} is mentioned in
\cite{Kiukkonenrichmobilephone2010,StachlPersonalityTraitsPredict2017}.

\paragraph{\textbf{(B) Let Users View Their Own Data}}
Only \cite{Kiukkonenrichmobilephone2010} discusses this aspect of privacy protection.
By letting the users see the data that is being collected, they can make a more informed
decision about sharing it.

\paragraph{\textbf{(C) Opt-out Option}}
The possibility of opting-out is only mentioned in \cite{WangStudentLifeAssessingMental2014}.
Especially after viewing their own data (see previous point), users might decide
that they no longer want to use the app or participate in the study.

\paragraph{\textbf{(D) Approval by Ethics Commission / Review Board}}
Psychological or medical studies typically require prior approval by an ethics commission or review board.
Three of the related works state that such approval was given for their studies
\cite{CanzianTrajectoriesDepressionUnobtrusive2015,FerreiraAWAREMobileContext2015,JayarajahLiveLabsBuildingInSitu2016}.
This aspect of privacy protection is more on a meta-level, as an ethics commission / review board
might check the other points mentioned in this privacy model.

\paragraph{\textbf{(E) Random Identifiers}}
When starting an app, often a login is required.
This poses the privacy risk of linking highly sensitive data with personal details, e.g.,
the user's Facebook account details if a Facebook account was used to log in.
Two related studies describe using random identifiers
\cite{WangStudentLifeAssessingMental2014,XuUnderstandingimpactpersonality2016}.
This point relies on the type of study being conducted.
Investigating the relationship between collected sensor data and,
for example, the number of Facebook friends, would probably require the
user to login via Facebook.
On a technical level for the Android system, an ID provided by the Google Play Services
proofed itself suitable as a random ID (cf.\ \cite{BeierleMobileSoft2018}).

\paragraph{\textbf{(F) Data Anonymization}}
This aspect is mentioned most commonly in the related work
\cite{Kiukkonenrichmobilephone2010,ChittaranjanWhoWhoBigFive2011,ChittaranjanMininglargescalesmartphone2013,LiKamWaMoodScopeBuildingMood2013,FerreiraAWAREMobileContext2015,JayarajahLiveLabsBuildingInSitu2016}.
If details are given, they usually describe how one-way hash functions are used
to obfuscate personally identifiable data like telephone numbers, Wifi SSIDs,
or Bluetooth addresses.

TYDR only stores clear text data where it is necessary for the research purpose.
Our context categorization from Section \ref{sec:cat-context} helps to analyze
why metadata will suffice in most cases.
Consider notifications for example.
Depending on the application, they might contain highly sensitive data, e.g.,
the message content of a messenger application.
The content of the notification is not relevant for our research purpose.
The app name that caused the notification however is, as one could easily
imagine a relationship between, e.g., the personality trait \emph{extraversion}
and the frequency of chat/messaging app notifications.

An additional point to consider regarding data anonymization is where the anonymization happens.
In \cite{JayarajahLiveLabsBuildingInSitu2016}, the authors describe how
the anonymization is taking place on the backend that the data is being sent to, before
being stored.
In TYDR, the anonymization process is taking place on the device itself, before storing
to the local device and before sending data to the backend.
So, even if our backend would be compromised,
the attacker would only be able to access data that is already anonymized.

\paragraph{\textbf{(G) Utilize Permission System}}
This point is specifically related to the Android permission system that was introduced
with Android 6.0 (cf.\ Section \ref{sec:cat-context}).
By itself, it can already make the users more aware of what data/sensor is being
accessed by an application.
The designers of an application still have influence over how they make use
of the system though.
Requesting all permissions at the first start of an app, e.g.,
gives the user little insight about what each permission is used for.
Instead, the app should request a permission at the point where it is needed
and explain to the user what the accessed data source is being used
for.

\paragraph{\textbf{(H) Secured Transfer}}
The point of having secured data transfer between mobile device
and backend is explicitly mentioned in
\cite{WangStudentLifeAssessingMental2014,FerreiraAWAREMobileContext2015}.
An alternative way is to only locally collect data and
ask users in a lab session to bring their phone and copy the data then.
Such an approach would severely limit the possible scope of a study.

\paragraph{\textbf{(I) Identifying Individual Users Without Linking to Their Collected Data}}

In psychological studies, it is common that users are compensated
with university course credit points, get paid to participate, or have the
chance to win money/vouchers in a raffle after study completion.
In order to contact the study participants,
contact information is needed, which
might contradict PM E.
In order to alleviate this concern, we developed a process
for identifying individual users without linking to their collected data on the backend \cite{BeierleMobileSoft2018}.
In short, the process consists of storing contact data separately from the collected smartphone data
and letting the app check the requirements for successful study completion, in our
case the daily completion of the PDD questionnaire.
This way, we can create incentives for users to install and use the app
while simultaneously preserving user privacy.

\section{Implementation of the Privacy Model \privmod in TYDR}
\label{sec:sys-arch}
\begin{wrapfigure}[29]{R}{0.4\textwidth}
	\centering
	\includegraphics[width=0.4\textwidth, trim = 0mm 0mm 0mm 0mm, clip=true]{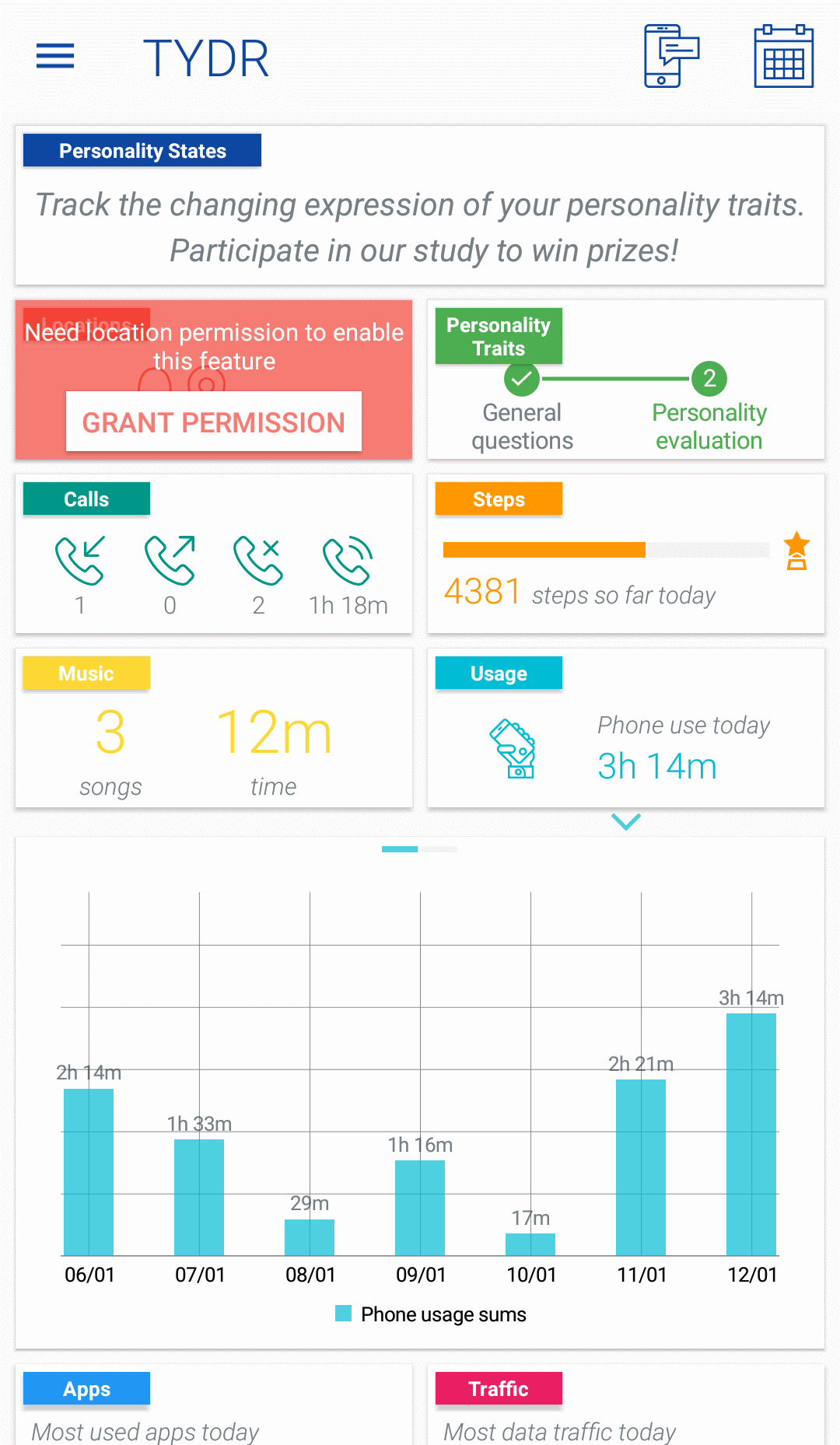}
	\caption{Main screen of TYDR.}
	\label{fig:screenshot}
\end{wrapfigure}
With TYDR, to the best of our knowledge, we are the first
to implement a privacy model
comprising all nine privacy measures listed in Section \ref{sec:privacy}.
The visualization of the data that is collected about the user is
TYDR's core feature (PM B).
The ethics commission of Technische Universit\"at Berlin approved
of using TYDR in a psychological study (PM D).

Figure \ref{fig:screenshot} shows TYDR's main screen and 
how it visualizes the collected data in a tile-based layout.
Each tile shows a daily summary of one data type.
By touching, a larger tile appears below with a weekly summary,
see for example the phone usage tile in the figure.
Users can opt-out via the contact form from the sidebar menu (PM C).
\begin{figure}[b]
	\centering
	\includegraphics[width=0.95\textwidth, trim = 0mm 0mm 0mm 0mm, clip=true]{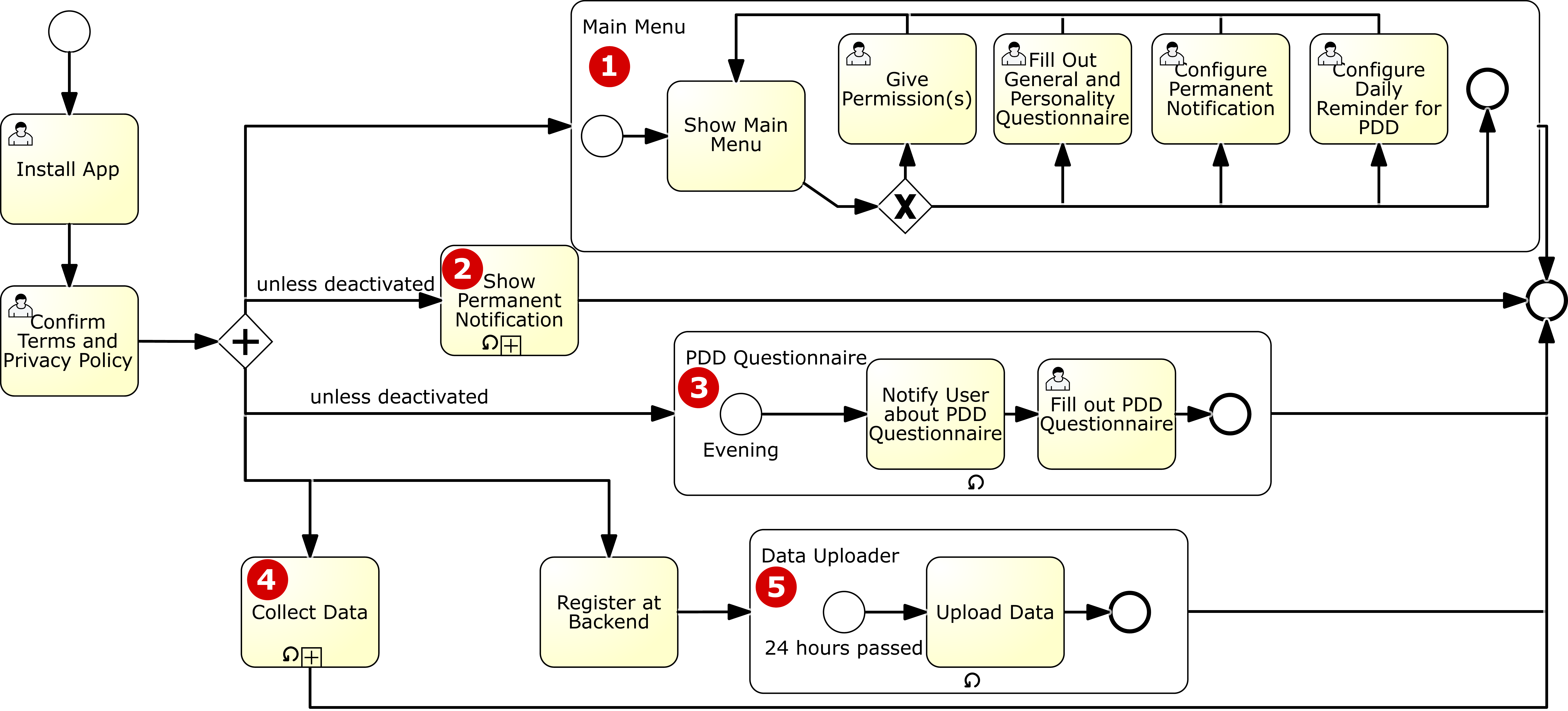}
	\caption{Process diagram for our smartphone sensor and usage data tracking app TYDR.}
	\label{fig:process}
\end{figure}
In Figure \ref{fig:process}, we show a diagram of the main processes in the TYDR app.
The person icon in a process signifies that the user is actively doing something.
All other processes are part of the app and do not require user interaction.
Starting TYDR for the first time, the user has to confirm the terms and the privacy policy (cf.\ PM A).
Only then the five processes of the app are started.
Note that there is no login process, the systems uses a random unique identifier (PM~E).

At the bottom of Figure \ref{fig:process}, we show that the app starts the data collection (Process 4).
The data collection engine already anonymizes the data before storing it (PM F).
The uploading via a secure connection (PM H)
is started after the app registered itself with the backend.
The upload process is repeated every 24 hours (Process 5).

The process at the top of the figure shows the main menu of TYDR (Process 1), also cf.\ Figure \ref{fig:screenshot}.
From here, the user can enable permissions
(cf.\ Table \ref{tbl:context} and Figure \ref{fig:screenshot}; PM G),
which influences the data collection.
The user can also fill out the general (demographic information)
and the personality traits questionnaire (\emph{Personality Traits} tile in Figure \ref{fig:screenshot}).
TYDR offers a permanent notification, displaying information on the lockscreen
and the notification bar (Process 2).
The data to be displayed can be configured by the user
via the second icon from the right in the top (Figure \ref{fig:screenshot}).
The tracking of personality states via the PDD questionnaire is designed to
be optional (Process 3).
Configuring the PDD questionnaire via the \emph{Personality States} tile
(Figure \ref{fig:screenshot}), the user can (de-)activate this feature.
In order to collect data labeled with personality states, we will conduct a study where
users commit to turning this feature on for a certain period of time.
The registration for this study takes into account PM I.

\section{Conclusion and Future Work}
\label{sec:concl-fw}

Context-aware applications
can potentially benefit from data
relating to the user's personality.
This includes rather static personality traits
and more dynamic personality states.
To be able to conduct a study on the relationship
between smartphone sensor and usage data and the user's
personality, we developed the Android app TYDR.
It tracks smartphone data and utilizes standardized
personality questionnaires.
TYDR tracks more types of data than existing related apps,
including metadata on notifications,
photos taken, and music listened to.

We developed a general context data model for smartphone applications,
highlighting the different kinds of interactions with the smartphone:
\emph{physical} conditions and activity, \emph{device} status and usage, \emph{core functions} usage, and \emph{app} usage.
We further developed the privacy model \privmod comprising nine proposed measures that can be taken to ensure user privacy
in apps related to mobile data collection.
On top of this, we presented the implementation of those nine measures for our
Android app TYDR.

Future work includes conducting the planned study to collect data for performing
data analysis to predict the user's personality from smartphone data.
This could comprise one prediction for personality traits and
daily predictions for personality states.
Based on our findings, we plan to develop a library
for the unobtrusive prediction of aspects of the user's personality
that can be utilized in context-aware applications.
The study results will have to show which permissions
will be necessary for such a library and
what categories of context will be the best predictors.
Regarding the privacy model, there are further questions to research, e.g.,
how to convey the privacy measures implemented to the user,
especially if they are not tech-savvy.

\section*{Acknowledgements}

This work was done in the context of
project DYNAMIC (\url{http://www.dynamic-project.de}) (grant No 01IS12056),
which is funded
as part of the Software Campus initiative by the German
Federal Ministry of Education and Research (BMBF).
We are grateful for the support provided by
Kai Grunert.

\bibliographystyle{elsarticle-harv}

\begin{thebibliography}{37}
\expandafter\ifx\csname natexlab\endcsname\relax\def\natexlab#1{#1}\fi
\providecommand{\url}[1]{\texttt{#1}}
\providecommand{\href}[2]{#2}
\providecommand{\path}[1]{#1}
\providecommand{\DOIprefix}{doi:}
\providecommand{\ArXivprefix}{arXiv:}
\providecommand{\URLprefix}{URL: }
\providecommand{\Pubmedprefix}{pmid:}
\providecommand{\doi}[1]{\href{http://dx.doi.org/#1}{\path{#1}}}
\providecommand{\Pubmed}[1]{\href{pmid:#1}{\path{#1}}}
\providecommand{\bibinfo}[2]{#2}
\ifx\xfnm\relax \def\xfnm[#1]{\unskip,\space#1}\fi
\bibitem[{Beierle(2018)}]{BeierleTrustCom2018}
\bibinfo{author}{Beierle, F.}, \bibinfo{year}{2018}.
\newblock \bibinfo{title}{{D}o {Y}ou {L}ike {W}hat {I} {L}ike? {S}imilarity
  {E}stimation in {P}roximity-based {M}obile {S}ocial {N}etworks}, in:
  \bibinfo{booktitle}{{P}roc. 2018 17th {IEEE} {I}nternational {C}onference
  {O}n {T}rust, {S}ecurity {A}nd {P}rivacy {I}n {C}omputing {A}nd
  {C}ommunications ({TrustCom})}, \bibinfo{publisher}{IEEE (to appear)}.
\bibitem[{Beierle et~al.(2015)Beierle, G{\"o}nd{\"o}r and
  K{\"u}pper}]{beierle-towards-2015}
\bibinfo{author}{Beierle, F.}, \bibinfo{author}{G{\"o}nd{\"o}r, S.},
  \bibinfo{author}{K{\"u}pper, A.}, \bibinfo{year}{2015}.
\newblock \bibinfo{title}{Towards a {{Three}}-tiered {{Social Graph}} in
  {{Decentralized Online Social Networks}}}, in: \bibinfo{booktitle}{Proc. 7th
  {{International Workshop}} on {{Hot Topics}} in {{Planet}}-Scale {{mObile
  Computing}} and {{Online Social neTworking}} ({{HotPOST}})},
  \bibinfo{publisher}{{ACM}}. pp. \bibinfo{pages}{1--6}.
\newblock \DOIprefix\doi{10.1145/2757513.2757517}.
\bibitem[{Beierle et~al.(2016)Beierle, Grunert, G{\"o}nd{\"o}r and
  K{\"u}pper}]{beierle-privacy-aware-2016}
\bibinfo{author}{Beierle, F.}, \bibinfo{author}{Grunert, K.},
  \bibinfo{author}{G{\"o}nd{\"o}r, S.}, \bibinfo{author}{K{\"u}pper, A.},
  \bibinfo{year}{2016}.
\newblock \bibinfo{title}{Privacy-aware {{Social Music Playlist Generation}}},
  in: \bibinfo{booktitle}{Proc. 2016 {{IEEE International Conference}} on
  {{Communications}} ({{ICC}})}, \bibinfo{publisher}{{IEEE}}. pp.
  \bibinfo{pages}{5650--5656}.
\newblock \DOIprefix\doi{10.1109/ICC.2016.7511602}.
\bibitem[{Beierle et~al.(2017)Beierle, Grunert, G{\"o}nd{\"o}r and
  Schl{\"u}ter}]{beierle-towards-2017}
\bibinfo{author}{Beierle, F.}, \bibinfo{author}{Grunert, K.},
  \bibinfo{author}{G{\"o}nd{\"o}r, S.}, \bibinfo{author}{Schl{\"u}ter, V.},
  \bibinfo{year}{2017}.
\newblock \bibinfo{title}{Towards {{Psychometrics}}-{{based Friend
  Recommendations}} in {{Social Networking Services}}}, in:
  \bibinfo{booktitle}{2017 {{IEEE International Conference}} on {{AI \& Mobile
  Services}} ({{AIMS}})}, \bibinfo{publisher}{{IEEE}}. pp.
  \bibinfo{pages}{105--108}.
\newblock \DOIprefix\doi{10.1109/AIMS.2017.22}.
\bibitem[{Beierle et~al.(2018)Beierle, Tran, Allemand, Neff, Schlee, Probst,
  Pryss and Zimmermann}]{BeierleMobileSoft2018}
\bibinfo{author}{Beierle, F.}, \bibinfo{author}{Tran, V.T.},
  \bibinfo{author}{Allemand, M.}, \bibinfo{author}{Neff, P.},
  \bibinfo{author}{Schlee, W.}, \bibinfo{author}{Probst, T.},
  \bibinfo{author}{Pryss, R.}, \bibinfo{author}{Zimmermann, J.},
  \bibinfo{year}{2018}.
\newblock \bibinfo{title}{{TYDR} -- {T}rack {Y}our {D}aily {R}outine. {A}ndroid
  {A}pp for {T}racking {S}martphone {S}ensor and {U}sage {D}ata}, in:
  \bibinfo{booktitle}{2018 {ACM/IEEE} 5th {I}nternational {C}onference on
  {M}obile {S}oftware {E}ngineering and {S}ystems {(MOBILESoft '18)}},
  \bibinfo{publisher}{ACM}. pp. \bibinfo{pages}{72--75}.
\newblock \DOIprefix\doi{10.1145/3197231.3197235}.
\bibitem[{Bogomolov et~al.(2014)Bogomolov, Lepri, Ferron, Pianesi and
  Pentland}]{BogomolovDailyStressRecognition2014}
\bibinfo{author}{Bogomolov, A.}, \bibinfo{author}{Lepri, B.},
  \bibinfo{author}{Ferron, M.}, \bibinfo{author}{Pianesi, F.},
  \bibinfo{author}{Pentland, A.S.}, \bibinfo{year}{2014}.
\newblock \bibinfo{title}{Daily {{Stress Recognition}} from {{Mobile Phone
  Data}}, {{Weather Conditions}} and {{Individual Traits}}}, in:
  \bibinfo{booktitle}{Proc. {{22nd ACM International Conference}} on
  {{Multimedia}}}, \bibinfo{publisher}{{ACM}}. pp. \bibinfo{pages}{477--486}.
\newblock \DOIprefix\doi{10.1145/2647868.2654933}.
\bibitem[{Butt and Phillips(2008)}]{ButtPersonalityselfreported2008}
\bibinfo{author}{Butt, S.}, \bibinfo{author}{Phillips, J.G.},
  \bibinfo{year}{2008}.
\newblock \bibinfo{title}{Personality and self reported mobile phone use}.
\newblock \bibinfo{journal}{Computers in Human Behavior} \bibinfo{volume}{24},
  \bibinfo{pages}{346--360}.
\bibitem[{Canzian and
  Musolesi(2015)}]{CanzianTrajectoriesDepressionUnobtrusive2015}
\bibinfo{author}{Canzian, L.}, \bibinfo{author}{Musolesi, M.},
  \bibinfo{year}{2015}.
\newblock \bibinfo{title}{Trajectories of {{Depression}}: {{Unobtrusive
  Monitoring}} of {{Depressive States}} by {{Means}} of {{Smartphone Mobility
  Traces Analysis}}}, in: \bibinfo{booktitle}{Proc. of the 2015 {{ACM
  International Joint Conference}} on {{Pervasive}} and {{Ubiquitous
  Computing}} ({UbiComp})}, \bibinfo{publisher}{{ACM}}. pp.
  \bibinfo{pages}{1293--1304}.
\newblock \DOIprefix\doi{10.1145/2750858.2805845}.
\bibitem[{Chittaranjan et~al.(2011)Chittaranjan, Blom and
  Gatica-Perez}]{ChittaranjanWhoWhoBigFive2011}
\bibinfo{author}{Chittaranjan, G.}, \bibinfo{author}{Blom, J.},
  \bibinfo{author}{Gatica-Perez, D.}, \bibinfo{year}{2011}.
\newblock \bibinfo{title}{Who's {{Who}} with {{Big}}-{{Five}}: {{Analyzing}}
  and {{Classifying Personality Traits}} with {{Smartphones}}}, in:
  \bibinfo{booktitle}{Proc. 2011 15th {{Annual International Symposium}} on
  {{Wearable Computers}}}, \bibinfo{publisher}{{IEEE}}. pp.
  \bibinfo{pages}{29--36}.
\newblock \DOIprefix\doi{10.1109/ISWC.2011.29}.
\bibitem[{Chittaranjan et~al.(2013)Chittaranjan, Blom and
  Gatica-Perez}]{ChittaranjanMininglargescalesmartphone2013}
\bibinfo{author}{Chittaranjan, G.}, \bibinfo{author}{Blom, J.},
  \bibinfo{author}{Gatica-Perez, D.}, \bibinfo{year}{2013}.
\newblock \bibinfo{title}{Mining large-scale smartphone data for personality
  studies}.
\newblock \bibinfo{journal}{Personal and Ubiquitous Computing}
  \bibinfo{volume}{17}, \bibinfo{pages}{433--450}.
\newblock \DOIprefix\doi{10.1007/s00779-011-0490-1}.
\bibitem[{Chorley et~al.(2015)Chorley, Whitaker and
  Allen}]{ChorleyPersonalitylocationbasedsocial2015}
\bibinfo{author}{Chorley, M.J.}, \bibinfo{author}{Whitaker, R.M.},
  \bibinfo{author}{Allen, S.M.}, \bibinfo{year}{2015}.
\newblock \bibinfo{title}{Personality and location-based social networks}.
\newblock \bibinfo{journal}{Computers in Human Behavior} \bibinfo{volume}{46},
  \bibinfo{pages}{45--56}.
\bibitem[{{de Montjoye} et~al.(2013){de Montjoye}, Quoidbach, Robic and
  Pentland}]{deMontjoyePredictingPersonalityUsing2013}
\bibinfo{author}{{de Montjoye}, Y.A.}, \bibinfo{author}{Quoidbach, J.},
  \bibinfo{author}{Robic, F.}, \bibinfo{author}{Pentland, A.},
  \bibinfo{year}{2013}.
\newblock \bibinfo{title}{Predicting {{Personality Using Novel Mobile
  Phone}}-{{Based Metrics}}}, in: \bibinfo{booktitle}{{{SBP}}},
  \bibinfo{publisher}{{Springer}}. pp. \bibinfo{pages}{48--55}.
\newblock \DOIprefix\doi{10.1007/978-3-642-37210-0_6}.
\bibitem[{Dey(2001)}]{dey-understanding-2001}
\bibinfo{author}{Dey, A.K.}, \bibinfo{year}{2001}.
\newblock \bibinfo{title}{Understanding and {Using} {Context}}.
\newblock \bibinfo{journal}{Personal Ubiquitous Comput.} \bibinfo{volume}{5},
  \bibinfo{pages}{4--7}.
\newblock \DOIprefix\doi{10.1007/s007790170019}.
\bibitem[{Ferreira et~al.(2015)Ferreira, Kostakos and
  Dey}]{FerreiraAWAREMobileContext2015}
\bibinfo{author}{Ferreira, D.}, \bibinfo{author}{Kostakos, V.},
  \bibinfo{author}{Dey, A.K.}, \bibinfo{year}{2015}.
\newblock \bibinfo{title}{{{AWARE}}: {{Mobile Context Instrumentation
  Framework}}}.
\newblock \bibinfo{journal}{Frontiers in ICT} \bibinfo{volume}{2}.
\bibitem[{Fleeson(2001)}]{FleesonStructureandProcessIntegratedView2001}
\bibinfo{author}{Fleeson, W.}, \bibinfo{year}{2001}.
\newblock \bibinfo{title}{Toward a {{Structure}}-and {{Process}}-{{Integrated
  View}} of {{Personality}}: {{Traits}} as {{Density Distributions}} of
  {{States}}}.
\newblock \bibinfo{journal}{Journal of Personality and Social Psychology}
  \bibinfo{volume}{80}, \bibinfo{pages}{1011--1027}.
\newblock \DOIprefix\doi{10.1037/0022-3514.80.6.1011}.
\bibitem[{Grover and Mark(2017)}]{GroverDigitalFootprintsPredicting2017}
\bibinfo{author}{Grover, T.}, \bibinfo{author}{Mark, G.}, \bibinfo{year}{2017}.
\newblock \bibinfo{title}{Digital {{Footprints}}: {{Predicting Personality}}
  from {{Temporal Patterns}} of {{Technology Use}}}, in:
  \bibinfo{booktitle}{Proc. 2017 {{ACM Intl. Joint Conference}} on
  {{Pervasive}} and {{Ubiquitous Computing}} and {{Proc.}} 2017 {{ACM Intl.
  Symposium}} on {{Wearable Computers}}}, \bibinfo{publisher}{{ACM}}. pp.
  \bibinfo{pages}{41--44}.
\bibitem[{Harari et~al.(2017)Harari, M{\"u}ller, Aung and
  Rentfrow}]{HarariSmartphonesensingmethods2017}
\bibinfo{author}{Harari, G.M.}, \bibinfo{author}{M{\"u}ller, S.R.},
  \bibinfo{author}{Aung, M.S.}, \bibinfo{author}{Rentfrow, P.J.},
  \bibinfo{year}{2017}.
\newblock \bibinfo{title}{Smartphone sensing methods for studying behavior in
  everyday life}.
\newblock \bibinfo{journal}{Current Opinion in Behavioral Sciences}
  \bibinfo{volume}{18}, \bibinfo{pages}{83--90}.
\newblock \DOIprefix\doi{10.1016/j.cobeha.2017.07.018}.
\bibitem[{Jayarajah et~al.(2016)Jayarajah, Balan, Radhakrishnan, Misra and
  Lee}]{JayarajahLiveLabsBuildingInSitu2016}
\bibinfo{author}{Jayarajah, K.}, \bibinfo{author}{Balan, R.K.},
  \bibinfo{author}{Radhakrishnan, M.}, \bibinfo{author}{Misra, A.},
  \bibinfo{author}{Lee, Y.}, \bibinfo{year}{2016}.
\newblock \bibinfo{title}{{{LiveLabs}}: {{Building In}}-{{Situ Mobile Sensing}}
  \& {{Behavioural Experimentation TestBeds}}}, in: \bibinfo{booktitle}{Proc.
  14th {{Annual International Conference}} on {{Mobile Systems}},
  {{Applications}}, and {{Services}}}, \bibinfo{publisher}{{ACM}}. pp.
  \bibinfo{pages}{1--15}.
\newblock \DOIprefix\doi{10.1145/2906388.2906400}.
\bibitem[{Karumur et~al.(2017)Karumur, Nguyen and
  Konstan}]{KarumurPersonalityUserPreferences2017}
\bibinfo{author}{Karumur, R.P.}, \bibinfo{author}{Nguyen, T.T.},
  \bibinfo{author}{Konstan, J.A.}, \bibinfo{year}{2017}.
\newblock \bibinfo{title}{Personality, {{User Preferences}} and {{Behavior}} in
  {{Recommender}} systems}.
\newblock \bibinfo{journal}{Information Systems Frontiers},
  \bibinfo{pages}{1--25}. \DOIprefix\doi{10.1007/s10796-017-9800-0}.
\bibitem[{Kiukkonen et~al.(2010)Kiukkonen, Blom, Dousse, Gatica-Perez and
  Laurila}]{Kiukkonenrichmobilephone2010}
\bibinfo{author}{Kiukkonen, N.}, \bibinfo{author}{Blom, J.},
  \bibinfo{author}{Dousse, O.}, \bibinfo{author}{Gatica-Perez, D.},
  \bibinfo{author}{Laurila, J.}, \bibinfo{year}{2010}.
\newblock \bibinfo{title}{Towards rich mobile phone datasets: {{Lausanne}} data
  collection campaign}, in: \bibinfo{booktitle}{Proc. {{ACM Intl}}. {{Conf}}.
  on {{Pervasive Computing}} ({{ICPS}})}.
\bibitem[{LiKamWa et~al.(2013)LiKamWa, Liu, Lane and
  Zhong}]{LiKamWaMoodScopeBuildingMood2013}
\bibinfo{author}{LiKamWa, R.}, \bibinfo{author}{Liu, Y.},
  \bibinfo{author}{Lane, N.D.}, \bibinfo{author}{Zhong, L.},
  \bibinfo{year}{2013}.
\newblock \bibinfo{title}{{{MoodScope}}: {{Building}} a {{Mood Sensor}} from
  {{Smartphone Usage Patterns}}}, in: \bibinfo{booktitle}{Proc. 11th {{Annual
  International Conference}} on {{Mobile Systems}}, {{Applications}}, and
  {{Services}}}, \bibinfo{publisher}{{ACM}}. pp. \bibinfo{pages}{389--402}.
\newblock \DOIprefix\doi{10.1145/2462456.2464449}.
\bibitem[{McCrae and John(1992)}]{mccrae-introduction-1992}
\bibinfo{author}{McCrae, R.R.}, \bibinfo{author}{John, O.P.},
  \bibinfo{year}{1992}.
\newblock \bibinfo{title}{An {{Introduction}} to the {{Five}}-{{Factor Model}}
  and {{Its Applications}}}.
\newblock \bibinfo{journal}{Journal of Personality} \bibinfo{volume}{60},
  \bibinfo{pages}{175--215}.
\bibitem[{{Michael J. Roche} et~al.(2014){Michael J. Roche}, {Aaron L. Pincus},
  {Amanda L. Rebar}, {David E. Conroy} and {Nilam
  Ram}}]{MichaelJ.RocheEnrichingPsychologicalAssessment2014}
\bibinfo{author}{{Michael J. Roche}}, \bibinfo{author}{{Aaron L. Pincus}},
  \bibinfo{author}{{Amanda L. Rebar}}, \bibinfo{author}{{David E. Conroy}},
  \bibinfo{author}{{Nilam Ram}}, \bibinfo{year}{2014}.
\newblock \bibinfo{title}{Enriching {{Psychological Assessment Using}} a
  {{Person}}-{{Specific Analysis}} of {{Interpersonal Processes}} in {{Daily
  Life}}}.
\newblock \bibinfo{journal}{Assessment} \bibinfo{volume}{21},
  \bibinfo{pages}{515--528}.
\newblock \DOIprefix\doi{10.1177/1073191114540320}.
\bibitem[{Mohr et~al.(2017)Mohr, Zhang and
  Schueller}]{MohrPersonalSensingUnderstanding2017}
\bibinfo{author}{Mohr, D.C.}, \bibinfo{author}{Zhang, M.},
  \bibinfo{author}{Schueller, S.M.}, \bibinfo{year}{2017}.
\newblock \bibinfo{title}{Personal {{Sensing}}: {{Understanding Mental Health
  Using Ubiquitous Sensors}} and {{Machine Learning}}}.
\newblock \bibinfo{journal}{Annual Review of Clinical Psychology}
  \bibinfo{volume}{13}, \bibinfo{pages}{23--47}.
\newblock \DOIprefix\doi{10.1146/annurev-clinpsy-032816-044949}.
\bibitem[{Myers et~al.(2010)Myers, Sen and
  Alexandrov}]{Myersmoderatingeffectpersonality2010}
\bibinfo{author}{Myers, S.D.}, \bibinfo{author}{Sen, S.},
  \bibinfo{author}{Alexandrov, A.}, \bibinfo{year}{2010}.
\newblock \bibinfo{title}{The moderating effect of personality traits on
  attitudes toward advertisements: A contingency framework}.
\newblock \bibinfo{journal}{Management \& Marketing} \bibinfo{volume}{5},
  \bibinfo{pages}{3--20}.
\bibitem[{Pryss et~al.(2018)Pryss, Probst, Schlee, Schobel, Langguth, Neff,
  Spiliopoulou and Reichert}]{PryssProspectivecrowdsensingretrospective2018}
\bibinfo{author}{Pryss, R.}, \bibinfo{author}{Probst, T.},
  \bibinfo{author}{Schlee, W.}, \bibinfo{author}{Schobel, J.},
  \bibinfo{author}{Langguth, B.}, \bibinfo{author}{Neff, P.},
  \bibinfo{author}{Spiliopoulou, M.}, \bibinfo{author}{Reichert, M.},
  \bibinfo{year}{2018}.
\newblock \bibinfo{title}{Prospective crowdsensing versus retrospective ratings
  of tinnitus variability and tinnitus\textendash{}stress associations based on
  the {{TrackYourTinnitus}} mobile platform}.
\newblock \bibinfo{journal}{International Journal of Data Science and
  Analytics}, \bibinfo{pages}{1--12}. \DOIprefix\doi{10.1007/s41060-018-0111-4}.
\bibitem[{Pryss et~al.(2015)Pryss, Reichert, Langguth and
  Schlee}]{PryssMobileCrowdSensing2015}
\bibinfo{author}{Pryss, R.}, \bibinfo{author}{Reichert, M.},
  \bibinfo{author}{Langguth, B.}, \bibinfo{author}{Schlee, W.},
  \bibinfo{year}{2015}.
\newblock \bibinfo{title}{Mobile {{Crowd Sensing Services}} for {{Tinnitus
  Assessment}}, {{Therapy}}, and {{Research}}}, in: \bibinfo{booktitle}{2015
  {{IEEE International Conference}} on Mobile {{Services}} ({{MS}})},
  \bibinfo{publisher}{{IEEE}}. pp. \bibinfo{pages}{352--359}.
\newblock \DOIprefix\doi{10.1109/MobServ.2015.55}.
\bibitem[{Rachuri et~al.(2010)Rachuri, Musolesi, Mascolo, Rentfrow, Longworth
  and Aucinas}]{RachuriEmotionSenseMobilePhones2010}
\bibinfo{author}{Rachuri, K.K.}, \bibinfo{author}{Musolesi, M.},
  \bibinfo{author}{Mascolo, C.}, \bibinfo{author}{Rentfrow, P.J.},
  \bibinfo{author}{Longworth, C.}, \bibinfo{author}{Aucinas, A.},
  \bibinfo{year}{2010}.
\newblock \bibinfo{title}{{{EmotionSense}}: {{A Mobile Phones Based Adaptive
  Platform}} for {{Experimental Social Psychology Research}}}, in:
  \bibinfo{booktitle}{Proc. 12th {{ACM Intl. Conference}} on {{Ubiquitous
  Computing}} ({UbiComp})}, \bibinfo{publisher}{{ACM}}. pp.
  \bibinfo{pages}{281--290}.
\newblock \DOIprefix\doi{10.1145/1864349.1864393}.
\bibitem[{Stachl et~al.(2017)Stachl, Hilbert, Au, Buschek, De~Luca, Bischl,
  Hussmann and B{\"u}hner}]{StachlPersonalityTraitsPredict2017}
\bibinfo{author}{Stachl, C.}, \bibinfo{author}{Hilbert, S.},
  \bibinfo{author}{Au, J.Q.}, \bibinfo{author}{Buschek, D.},
  \bibinfo{author}{De~Luca, A.}, \bibinfo{author}{Bischl, B.},
  \bibinfo{author}{Hussmann, H.}, \bibinfo{author}{B{\"u}hner, M.},
  \bibinfo{year}{2017}.
\newblock \bibinfo{title}{Personality {{Traits Predict Smartphone Usage}}}.
\newblock \bibinfo{journal}{European Journal of Personality}
  \bibinfo{volume}{31}, \bibinfo{pages}{701--722}.
\newblock \DOIprefix\doi{10.1002/per.2113}.
\bibitem[{Wang et~al.(2014)Wang, Chen, Chen, Li, Harari, Tignor, Zhou, Ben-Zeev
  and Campbell}]{WangStudentLifeAssessingMental2014}
\bibinfo{author}{Wang, R.}, \bibinfo{author}{Chen, F.}, \bibinfo{author}{Chen,
  Z.}, \bibinfo{author}{Li, T.}, \bibinfo{author}{Harari, G.},
  \bibinfo{author}{Tignor, S.}, \bibinfo{author}{Zhou, X.},
  \bibinfo{author}{Ben-Zeev, D.}, \bibinfo{author}{Campbell, A.T.},
  \bibinfo{year}{2014}.
\newblock \bibinfo{title}{{{StudentLife}}: {{Assessing Mental Health}},
  {{Academic Performance}} and {{Behavioral Trends}} of {{College Students
  Using Smartphones}}}, in: \bibinfo{booktitle}{Proc. 2014 {{ACM International
  Joint Conference}} on {{Pervasive}} and {{Ubiquitous Computing}}
  ({UbiComp})}, \bibinfo{publisher}{{ACM}}. pp. \bibinfo{pages}{3--14}.
\newblock \DOIprefix\doi{10.1145/2632048.2632054}.
\bibitem[{Wang et~al.(2017)Wang, Chen, Chen, Li, Harari, Tignor, Zhou, Ben-Zeev
  and Campbell}]{WangStudentLifeUsingSmartphones2017}
\bibinfo{author}{Wang, R.}, \bibinfo{author}{Chen, F.}, \bibinfo{author}{Chen,
  Z.}, \bibinfo{author}{Li, T.}, \bibinfo{author}{Harari, G.},
  \bibinfo{author}{Tignor, S.}, \bibinfo{author}{Zhou, X.},
  \bibinfo{author}{Ben-Zeev, D.}, \bibinfo{author}{Campbell, A.T.},
  \bibinfo{year}{2017}.
\newblock \bibinfo{title}{{{StudentLife}}: {{Using Smartphones}} to {{Assess
  Mental Health}} and {{Academic Performance}} of {{College Students}}}, in:
  \bibinfo{booktitle}{Mobile {{Health}}}. \bibinfo{publisher}{{Springer}}, pp.
  \bibinfo{pages}{7--33}.
\newblock \DOIprefix\doi{10.1007/978-3-319-51394-2_2}.
\bibitem[{Wang et~al.(2015)Wang, Harari, Hao, Zhou and
  Campbell}]{WangSmartGPAHowSmartphones2015}
\bibinfo{author}{Wang, R.}, \bibinfo{author}{Harari, G.}, \bibinfo{author}{Hao,
  P.}, \bibinfo{author}{Zhou, X.}, \bibinfo{author}{Campbell, A.T.},
  \bibinfo{year}{2015}.
\newblock \bibinfo{title}{{{SmartGPA}}: {{How Smartphones Can Assess}} and
  {{Predict Academic Performance}} of {{College Students}}}, in:
  \bibinfo{booktitle}{Proc. 2015 {{ACM International Joint Conference}} on
  {{Pervasive}} and {{Ubiquitous Computing}} ({UbiComp})},
  \bibinfo{publisher}{{ACM}}. pp. \bibinfo{pages}{295--306}.
\newblock \DOIprefix\doi{10.1145/2750858.2804251}.
\bibitem[{Xiong et~al.(2016)Xiong, Huang, Barnes and
  Gerber}]{XiongSensusCrossplatformGeneralpurpose2016}
\bibinfo{author}{Xiong, H.}, \bibinfo{author}{Huang, Y.},
  \bibinfo{author}{Barnes, L.E.}, \bibinfo{author}{Gerber, M.S.},
  \bibinfo{year}{2016}.
\newblock \bibinfo{title}{Sensus: {{A Cross}}-platform, {{General}}-purpose
  {{System}} for {{Mobile Crowdsensing}} in {{Human}}-subject {{Studies}}}, in:
  \bibinfo{booktitle}{Proc. 2016 {{ACM International Joint Conference}} on
  {{Pervasive}} and {{Ubiquitous Computing}} ({UbiComp})},
  \bibinfo{publisher}{{ACM}}. pp. \bibinfo{pages}{415--426}.
\newblock \DOIprefix\doi{10.1145/2971648.2971711}.
\bibitem[{Xu et~al.(2016)Xu, Frey, Fleisch and
  Ilic}]{XuUnderstandingimpactpersonality2016}
\bibinfo{author}{Xu, R.}, \bibinfo{author}{Frey, R.M.},
  \bibinfo{author}{Fleisch, E.}, \bibinfo{author}{Ilic, A.},
  \bibinfo{year}{2016}.
\newblock \bibinfo{title}{Understanding the impact of personality traits on
  mobile app adoption \textendash{} {{Insights}} from a large-scale field
  study}.
\newblock \bibinfo{journal}{Computers in Human Behavior} \bibinfo{volume}{62},
  \bibinfo{pages}{244--256}.
\newblock \DOIprefix\doi{10.1016/j.chb.2016.04.011}.
\bibitem[{Yurur et~al.(2014)Yurur, Liu, Sheng, Leung, Moreno and
  Leung}]{yurur-context-awareness-2014}
\bibinfo{author}{Yurur, O.}, \bibinfo{author}{Liu, C.}, \bibinfo{author}{Sheng,
  Z.}, \bibinfo{author}{Leung, V.}, \bibinfo{author}{Moreno, W.},
  \bibinfo{author}{Leung, K.}, \bibinfo{year}{2014}.
\newblock \bibinfo{title}{Context-{Awareness} for {Mobile} {Sensing}: {A}
  {Survey} and {Future} {Directions}}.
\newblock \bibinfo{journal}{IEEE Communications Surveys Tutorials}
  \bibinfo{volume}{18}, \bibinfo{pages}{1--28}.
\newblock \DOIprefix\doi{10.1109/COMST.2014.2381246}.
\bibitem[{Zhou and Lu(2011)}]{ZhouEffectsPersonalityTraits2011}
\bibinfo{author}{Zhou, T.}, \bibinfo{author}{Lu, Y.}, \bibinfo{year}{2011}.
\newblock \bibinfo{title}{The {{Effects}} of {{Personality Traits}} on {{User
  Acceptance}} of {{Mobile Commerce}}}.
\newblock \bibinfo{journal}{International Journal of Human\textendash{}Computer
  Interaction} \bibinfo{volume}{27}, \bibinfo{pages}{545--561}.
\newblock \DOIprefix\doi{10.1080/10447318.2011.555298}.
\bibitem[{Zimmermann et~al.(2018)Zimmermann, Woods, Ritter, Happel, Masuhr,
  Jaeger, Spitzer and Wright}]{ZimmermannIntegratingstructuredynamics}
\bibinfo{author}{Zimmermann, J.}, \bibinfo{author}{Woods, W.C.},
  \bibinfo{author}{Ritter, S.}, \bibinfo{author}{Happel, M.},
  \bibinfo{author}{Masuhr, O.}, \bibinfo{author}{Jaeger, U.},
  \bibinfo{author}{Spitzer, C.}, \bibinfo{author}{Wright, A.G.C.},
  \bibinfo{year}{2018}.
\newblock \bibinfo{title}{Integrating structure and dynamics in personality
  assessment: {{First}} steps toward the development and validation of a
  {{Personality Dynamics Diary}}}.
\newblock \bibinfo{journal}{Psychological Assessment (in press)}.

\end{thebibliography}

\clearpage

\normalMode

\end{document}